\documentclass[11pt,a4paper]{article}
\date{}

\usepackage{latexsym}
\usepackage{amsfonts}
\usepackage[all]{xy}
\usepackage{amssymb, mathrsfs, amsfonts, amsmath}
\usepackage{amsbsy}

\newtheorem{thm}{Theorem}[section]
 \newtheorem{cor}[thm]{Corollary}
 \newtheorem{lem}[thm]{Lemma}
 \newtheorem{prop}[thm]{Proposition}
  \newtheorem{defn}[thm]{Definition}
 \newtheorem{rem}[thm]{Remark}
 \numberwithin{equation}{section}

\begin{document}

\author{Tulkin H. Rasulov, Mukhiddin I. Muminov and Mahir Hasanov}
\title{ \bf On the Spectrum of a Model Operator in Fock Space} \maketitle
\begin{abstract}
A model operator $H$ associated to a system describing four
particles in interaction, without conservation of the number of
particles, is considered. We  describe the essential spectrum of
$H$ by the spectrum of the channel operators and prove the
Hunziker-van Winter-Zhislin (HWZ) theorem for the operator $H.$ We
also give some variational principles  for boundaries of the
essential spectrum and interior eigenvalues.
\end{abstract}

 \vspace{0.1cm}


\section{INTRODUCTION}

Spectral properties of multi-particle continuous Schr\"{o}dinger
operators are sufficiently well studied in \cite{Fad-Merk}. As is
well-known, the theorem on the location of the essential spectrum
of multi-particle Hamiltonians was named the HWZ theorem in
\cite{C-F-K-S,RS-4} to the honor of Hunziker \cite{Hunziker}, van
Winter \cite{Van Winter} and Zhislin \cite{Zhislin}. A lattice
analogue of this theorem for the four-particle Schr\"{o}dinger
operator is proved in \cite{ALA,MM}.

The effective  description of the location of the essential
spectrum of electromagnetic Schr\"{o}dinger operators on ${\bf
R}^N$  is obtained  in \cite{Rab}. The well-known methods for the
investigation of the location of essential spectra of
Schr\"{o}dinger operators are Weyl criterion for the one particle
problem and the HWZ theorem for multiparticle problems, the modern
proof of which is based on the Ruelle-Simon partition of unity. In
\cite{Rab-Roch} by means of the limit operators method the
essential spectrum of discrete Schr\"{o}dinger operators on
lattice ${\bf Z}^N$ is studied. This method has been applied by
one of the authors to describe the essential spectrum of
continuous electromagnetic Schr\"{o}dinger operators, square-root
Klein-Gordon operators and Dirac operators under quite weak
assumptions on the behavior of the magnetic and electric potential
at infinity.

The systems considered above have a fixed number of
quasi-particles. In statistical physics \cite{Min-Sp}, solid-state
physics \cite{Mog} and the theory of quantum fields \cite{Frid},
one considers systems, where the number of quasi-particles is
bounded, but not fixed. The study of these systems usually is
reduced to the study of the spectral properties of self-adjoint
operators, acting in the {\it cut subspace} ${\cal H}^{(N)}$ of
Fock space, consisting of $n\leq N$ particles
\cite{Frid,Min-Sp,Mog,SSZ,Zhu-Minl}.

In \cite{SSZ} geometric and commutator techniques have been
developed in order to find the location of the spectrum and to
prove absence of singular continuous spectrum for Hamiltonians
without conservation of the particle number. The model operators
acting in ${\cal H}^{(3)}$ were well studied in
\cite{ALR-1,ALR-2,LR-1,LR-2,R-1,Yod-Mum}.

In the present paper we consider a model operator $H$ associated
to a system describing four particles in interaction, without
conservation of the number of particles, acting in ${\cal
H}^{(4)}.$ For the study of location of the essential spectrum of
$H$ we introduce the channel operators and prove that the
essential spectrum of $H$ is the union of spectra of channel
operators. The channel operators have a more simple structure than
$H.$ The two-particle, three-particle and four-particle branches
of the essential spectrum of $H$ are singled out. We also prove
the HWZ theorem on the location of the essential spectrum of $H.$
A variational approach to find boundaries of essential spectrum
and some interior eigenvalues is given at the end of the paper.

The plan of the present paper is as follows.

Section 1 is an introduction to the whole work. In Section 2 the
model operator $H$ is described as a bounded self-adjoint operator
in ${\cal H}^{(4)}$ and the main results of the present paper are
formulated. In Section 3 we study  spectrum of channel operators
by the spectrum of corresponding families of operators. In Section
4 we obtain analogue of the Faddeev-Yakubovskii type system of
integral equations for the eigenvectors of $H.$ Section 5 is
devoted to the proof of the main results of the present paper
(Theorems \ref{ess of H} and \ref{HWZ}). In Section 6 we apply
some results from classical variational theory and the variational
theory of the spectrum of operator pencils to the model operator
$H.$

Throughout the present paper we adopt the following convention:
Denote by  ${\bf T}^\nu$ the $\nu$-dimensional torus, the cube
$(-\pi,\pi]^\nu$ with appropriately identified sides. The torus
${\bf T}^\nu$ will always be considered as an abelian group with
respect to the addition and multiplication by real numbers
regarded as operations on the $\nu$-dimentional space ${\bf
R}^\nu$ modulo $(2 \pi {\bf Z})^\nu.$

\section{THE MODEL OPERATOR AND STATEMENTS OF THE MAIN RESULTS}

Let us introduce some notations used in this work. Let ${\bf C}$
be the field of complex numbers, $({\bf T}^\nu)^n,\,n=1,2,3$ be
the Cartesian $n$th power of ${\bf T}^\nu$ and $L_2(({\bf
T}^\nu)^n),\,n=1,2,3$ be the Hilbert space of square-integrable
(complex) functions defined on $({\bf T}^\nu)^n,\,n=1,2,3.$

Denote
\begin{equation*}
{\cal H}_0={\bf C},\,{\cal H}_1=L_2({\bf T}^\nu),\,{\cal
H}_2=L_2(({\bf T}^\nu)^2),\,{\cal H}_3=L_2(({\bf T}^\nu)^3),
\end{equation*}
\begin{equation*}
{\cal H}^{(n,m)}=\bigoplus_{i=n}^m {\cal H}_i,\, 0\leq n<m \leq 3.
\end{equation*}

The Hilbert space ${\cal H}^{(4)}\equiv {\cal H}^{(0,3)}$ is
called {\it "four-particle cut subspace"} of Fock space.

Let the model operator $H$ act in the Hilbert space ${\cal
H}^{(0,3)}$ as a matrix operator
\begin{equation}\label{main operator}
H=\left( \begin{array}{cccc}
H_{00} & H_{01} & 0 & 0\\
H_{10} & H_{11} & H_{12} & 0\\
0 & H_{21} & H_{22} & H_{23}\\
0 & 0 & H_{32} & H_{33}\\
\end{array}
\right),
\end{equation}
where its components $H_{ij}: {\cal{H}}_j\to {\cal{H}}_i,\,\
i,j=0,1,2,3$ are defined by the rule
\begin{equation*}
(H_{00}f_0)_0=w_0f_0,\,\, (H_{01}f_1)_0=\int\limits_{{\bf T}^\nu}
v_1(s)f_1(s)ds,\,\,  (H_{10}f_0)_1(p)=v_1(p)f_0,
\end{equation*}
\begin{equation*}
(H_{11}f_1)_1(p)=w_1(p)f_1(p),\quad
(H_{12}f_2)_1(p)=\int\limits_{{\bf T}^\nu} v_2(s)f_2(p, s)ds,
\end{equation*}
\begin{equation*}
(H_{21}f_1)_2(p, q)=v_2(q)f_1(p), \,\,
H_{22}=H_{22}^0-V_{21}-V_{22},
\end{equation*}
\begin{equation*}
(H_{22}^0f_2)_2(p, q)=w_2(p, q)f_2(p, q),\, (V_{21}f_2)_2(p,
q)=v_{21}(p)\int\limits_{{\bf T}^\nu} v_{21}(s)f_2(s, q)ds,
\end{equation*}
\begin{equation*}
(V_{22}f_2)_2(p, q)=v_{22}(q)\int\limits_{{\bf T}^\nu}
v_{22}(s)f_2(p, s)ds,\, (H_{23}f_3)_2(p, q)=\int\limits_{{\bf
T}^\nu} v_3(s)f_3(p, q, s)ds,
\end{equation*}
\begin{equation*}
(H_{32}f_2)_3(p, q, t)= v_3(t)f_2(p, q),\,\, (H_{33}f_3)_3(p, q,
t)=w_3(p, q, t)f_3(p, q, t).
\end{equation*}

Here $f_i\in {\cal{H}}_i,\,i=0,1,2,3,\,w_0$ is a real number,
$v_i(\cdot),\,i=1,2,3,$ $v_{2j}(\cdot),\,j=1,2,\,w_1(\cdot)$ are
real-valued continuous functions on ${\bf T}^\nu$ and $w_2(\cdot,
\cdot)$ resp. $w_3(\cdot, \cdot, \cdot)$ is a real-valued
continuous function on $({\bf T}^\nu)^2$ resp. $({\bf T}^\nu)^3.$

Under these assumptions the operator $H$ is bounded and
self-adjoint in ${\cal H}^{(0,3)}.$

We remark that the operators $H_{01},$ $H_{12}$ and $H_{23}$ resp.
$H_{10},$ $H_{21}$ and $H_{32}$ defined in the Fock space are
called annihilation resp. creation operators.

To formulate the main results of the present paper we introduce
the following channel operators $H_{n},\,n=1,3$ resp. $H_2$ acting
in ${\cal{H}}^{(2,3)}$  resp. ${\cal{H}}^{(1,3)}$ by the following
formula
\begin{equation*}
H_1=\left( \begin{array}{cc}
H_{22}^0-V_{21} & H_{23}\\
H_{32} & H_{33} \\
\end{array}
\right), \quad H_3=\left( \begin{array}{cc}
H_{22}^0 & H_{23}\\
H_{32} & H_{33}\\
\end{array}
\right)
\end{equation*}
resp.
\begin{equation*}
H_2=\left( \begin{array}{ccc}
H_{11} & H_{12} & 0\\
H_{21} & H_{22}^0-V_{22} & H_{23}\\
0 & H_{32} & H_{33}\\
\end{array}
\right).
\end{equation*}

Now we give the main results of the paper (for the proof see
Section 5).

The essential spectrum of the operator $H$ can be precisely
described as well as in the following

\begin{thm}\label{ess of H} The essential spectrum
$\sigma_{ess}(H)$ of the operator $H$ is the union of spectra of
channel operators $H_1,\,H_2$ and $H_3,$  i. e., the equality
$$\sigma_{ess}(H)= \bigcup_{n=1}^3 \sigma(H_n)$$
holds, where $\sigma(H_n),\,n=1,2,3$ stands the spectrum of the
operator $H_n,\,n=1,2,3.$
\end{thm}

The following theorem shows that the least element of the
essential spectrum of $H$ belongs to the spectrum of channel
operator $H_1$ or $H_2.$

\begin{thm}\label{HWZ} (a HWZ theorem). The following equality
\begin{equation*}
\min \sigma_{ess}(H)=\min\{\min\sigma (H_1),\,\min\sigma (H_2)\}
\end{equation*}
holds.
\end{thm}

\section{THE SPECTRUM OF THE CHANNEL OPERATORS}

In this section we describe the spectrum of the channel operators
$H_n,\,n=1,2$ resp. $H_3$ by the spectrum of the family of
operators $h_n(p),\,p\in {\bf T}^\nu,\,n=1,2$ resp. $h_3(p,q),\,p,
q\in {\bf T}^\nu$ defined below.

First we consider the operator $H_3,$ which commutes with any
multiplication operator $U_\alpha^{(3)}$ by the bounded function
$\alpha(\cdot,\cdot)$ on $({\bf T}^\nu)^2$
\begin{equation*}
U_\alpha^{(3)}\left( \begin{array}{cc}
g_2(p, q)\\
g_3(p, q, t)\\
\end{array}
\right)=\left( \begin{array}{cc}
\alpha(p, q) g_2(p, q)\\
\alpha(p, q) g_3(p, q, t)\\
\end{array}
\right),\,\left( \begin{array}{cc}
g_2\\
g_3\\
\end{array}
\right) \in {\cal H}^{(2,3)}.
\end{equation*}

Therefore the decomposition of the space ${\cal H}^{(2,3)}$ into
the direct integral
\begin{equation*}
{\cal H}^{(0,1)}= \int\limits_{({\bf T}^\nu)^2} \oplus \,{\cal
H}^{(2,3)}dpdq
\end{equation*}
yields the decomposition into the direct integral
\begin{equation}\label{decom H_3}
H_3= \int\limits_{({\bf T}^\nu)^2} \oplus\, h_3(p, q)dpdq,
\end{equation}
where a family of the generalized Friedrichs models $h_3(p,
q),\,p, q\in {\bf T}^\nu$ acts in ${\cal H}^{(0,1)}$ as
\begin{equation*}
h_3(p, q)=\left( \begin{array}{cc}
h_{00}^{(3)}(p, q) & h_{01}^{(3)}\\
h_{10}^{(3)} & h_{11}^{(3)}(p, q)\\
\end{array}
\right).
\end{equation*}
Here
\begin{equation*}
(h_{00}^{(3)}(p, q)f_0)_0=w_2(p, q)f_0,\quad
(h_{01}^{(3)}f_1)_0=\int\limits_{{\bf T}^\nu} v_3(s)f_1(s)ds,
\end{equation*}
\begin{equation*}
(h_{10}^{(3)}f_0)_1(t)=v_3(t)f_0,\quad (h_{11}^{(3)}(p,
q)f_1)_1(t)=w_3(p, q, t)f_1(t).
\end{equation*}

In analogy with the operator $H_3$ one can give the decomposition
\begin{equation}\label{decom H_i}
H_n= \int\limits_{{\bf T}^\nu} \oplus \,h_n(p)dp,\,n=1,2,
\end{equation}
where a family of the operators $h_1(p),\,p\in {\bf T}^\nu$ resp.
$h_2(p),\,p\in {\bf T}^\nu$ acts in ${\cal H}^{(1,2)}$ resp.
${\cal H}^{(0,2)}$ as
\begin{equation*}
h_1(p)=\left( \begin{array}{cc}
h_{11}^{(1)}(p) & h_{12}^{(1)}\\
h_{21}^{(1)} & h_{22}^{(1)}(p)\\
\end{array}
\right)\quad\mbox{resp.}\quad h_2(p)=\left( \begin{array}{ccc}
h_{00}^{(2)}(p) & h_{01}^{(2)} & 0\\
h_{10}^{(2)} & h_{11}^{(2)}(p) & h_{12}^{(1)}\\
0 & h_{21}^{(1)} & h_{22}^{(1)}(p)\\
\end{array}
\right)
\end{equation*}
with the entries
\begin{equation*}
(h_{11}^{(1)}(p)f_1)_1(q)=w_2(p,
q)f_1(q)-v_{21}(q)\int\limits_{{\bf T}^\nu} v_{21}(s)f_1(s)ds,\,
(h_{12}^{(1)}f_2)_1(q)=\int\limits_{{\bf T}^\nu} v_3(s)f_2(q,
s)ds,
\end{equation*}
\begin{equation*}
(h_{21}^{(1)}f_1)_2(q, t)=v_3(t)f_1(q),\quad
(h_{22}^{(1)}(p)f_2)_2(q, t)=w_3(p, q, t)f_2(q, t),
\end{equation*}
\begin{equation*}
(h_{00}^{(2)}(p)f_0)_0=w_1(p)f_0,\,
(h_{01}^{(2)}f_1)_0=\int\limits_{{\bf T}^\nu} v_2(s)f_1(s)ds,\,
(h_{10}^{(2)}f_0)_1(q)=v_2(q)f_0,
\end{equation*}
\begin{equation*}
(h_{11}^{(2)}(p)f_1)_1(q)=w_2(p,
q)f_1(q)-v_{22}(q)\int\limits_{{\bf T}^\nu} v_{22}(s)f_1(s)ds.
\end{equation*}

Let us introduce the notations
\begin{equation*}
m= \min_{p, q, t\in {\bf T}^\nu} w_3(p, q, t),\,\, M= \max_{p, q,
t\in {\bf T}^\nu} w_3(p, q, t),
\end{equation*}
\begin{equation*}
\sigma_{four}(H_n)=[m; M],\,n=1,2,3,
\end{equation*}
\begin{equation*}
\sigma_{three}(H_n)=\bigcup_{p, q\in {\bf T}^\nu}
\sigma_{disc}(h_3(p, q)),\,n=1,2,3,
\end{equation*}
\begin{equation*}
\sigma_{two}(H_n)=\bigcup\limits_{p\in {\bf T}^\nu}
\sigma_{disc}(h_n(p)),\,n=1,2.
\end{equation*}

The spectrum of the operators $H_n,\,n=1,2,3$ can be precisely
described as well as in the following

\begin{thm}\label{chanl spectr} The following equalities hold:\\
(i) $\sigma(H_1)=\sigma_{two}(H_1) \cup \sigma_{three}(H_1) \cup
\sigma_{four}(H_1);$\\
(ii) $\sigma(H_2)=\sigma_{two}(H_2) \cup \sigma_{three}(H_2) \cup \sigma_{four}(H_2);$\\
(iii) $\sigma(H_3)=\sigma_{three}(H_3) \cup \sigma_{four}(H_3).$
\end{thm}

Before proving the Theorem \ref{chanl spectr} we introduce a new
subsets of the essential spectrum of H.

\begin{defn}\label{branches} The sets
$\sigma_{two}(H)=\sigma_{two}(H_1)\cup \sigma_{two}(H_2),$
$\sigma_{three}(H)=\sigma_{three}(H_3)$ and
$\sigma_{four}(H)=\sigma_{four}(H_3)$ are called two-particle,
three-particle and four-particle branches of the essential
spectrum of $H,$ respectively.
\end{defn}

We starts the proof of the Theorem \ref{chanl spectr} with the
following auxiliary statements.

Let the operator $h_3^0(p,q),\,p,q\in {\bf T}^\nu$ acts in ${\cal
H}^{(0,1)}$ as
\begin{equation*}
h_3^0(p,q)=\left( \begin{array}{cc}
0 & 0\\
0 & h_{11}^{(3)}(p,q)\\
\end{array}
\right),\,p,q\in {\bf T}^\nu.
\end{equation*}

The perturbation $h_3(p,q)-h_3^0(p,q),\,p,q\in {\bf T}^\nu$ of the
operator $h_3^0(p,q),\,p,q\in {\bf T}^\nu$ is a self-adjoint
operator of rank 2. Therefore in accordance with the invariance of
the essential spectrum under finite rank perturbations the
essential spectrum $\sigma_{ess}(h_3(p,q))$ of $h_3(p,q),\,p,q\in
{\bf T}^\nu$ fills the following interval on the real axis:
\begin{equation*}
\sigma_{ess}(h_3(p,q))=[m_3(p,q);M_3(p,q)],
\end{equation*}
where the numbers  $m_3(p,q)$ and $M_3(p,q)$ are defined by
\begin{equation*}
m_3 (p,q)= \min_{t\in {\bf T}^\nu} w_3(p,q,t),\quad M_3(p,q)=
\max_{t\in {\bf T}^\nu} w_3(p,q,t).
\end{equation*}

\begin{rem} We remark that for some $p,q\in {\bf T}^\nu$
the essential spectrum of $h_3(p,q)$ may degenerate to the set
consisting of the unique point $\{m_3(p,q)\}$ and hence we cannot
state that the essential spectrum of $h_3(p,q)$ is absolutely
continuous for any $p,q \in {\bf T}^\nu.$ For example, this is the
case if the function $w_3(\cdot, \cdot, \cdot)$ is of the form
\begin{equation*}
w_3(p, q, t)=\varepsilon(p)+\varepsilon(q+t)+\varepsilon(t),
\end{equation*}
where $p=q=(\underbrace{\pi,\ldots,\pi}\limits_{\nu})\in {\bf
T}^\nu$ and
\begin{equation*}
\varepsilon(t)=\nu-\sum\limits_{i=1}^{\nu}\cos t_i,\, t=(t_1,
t_2,\ldots,t_{\nu})\in {\bf T}^\nu.
\end{equation*}
\end{rem}

For any fixing $p,q\in {\bf T}^\nu$ we define an analytic function $
\Delta_3(p,q\,; \cdot)$ (the Fredholm determinant associated with
the operator $h_3(p,q),\,p,q\in {\bf T}^\nu$) in ${\bf C}\setminus
\sigma_{ess} (h_3(p,q))$ by
\begin{equation*}
\Delta_3(p,q\,; z)=w_2(p,q)-z-\int\limits_{{\bf T}^\nu}\frac{
v_3^2(s)ds}{w_3(p,q,s)-z}.
\end{equation*}

The following lemma established a connection between of
eigenvalues of $h_3(p,q),\,p,q\in {\bf T}^\nu$ and the zeroes of
the function $ \Delta_3(p,q\,; \cdot),\,p,q\in {\bf T}^\nu.$

\begin{lem}\label{Delta_3} For any fixing $p,q \in {\bf T}^\nu$ the number $z \in {\bf
C} \setminus \sigma_{ess} (h_3(p,q))$ is an eigenvalue of the
operator $h_3(p,q),\,p,q\in {\bf T}^\nu$ if and only if $
\Delta_3(p,q\,; z)=0.$
\end{lem}

{\bf Proof.} "Only If Part." Let for any fixing $p,q \in {\bf
T}^\nu$ the number $z \in {\bf C}\setminus \sigma_{ess}
(h_3(p,q))$ be an eigenvalue of the operator $h_3(p,q),\,p,q \in
{\bf T}^\nu$ and $f=(f_0,f_1)\in {\cal H}^{(0,1)}$ be the
corresponding eigenvector, i. e., the equation $h_3(p,q)f=zf$ or
the system of equations
\begin{equation}\label{sistem h_3}
\left \lbrace
\begin{array}{ll}
(w_2(p,q)-z)f_0+ \int\limits_{{\bf T}^\nu} v_3(s)f_1(s)ds=0\\
v_3(t)f_0+(w_3(p,q,t)-z)f_1(t)=0
\end{array} \right.
\end{equation}
has a nontrivial solution $f=(f_0,f_1)\in {\cal H}^{(0,1)}.$

Since $z \in {\bf C}\setminus \sigma_{ess} (h_3(p,q))$ from the
second equation of the system (\ref{sistem h_3}) we  find
\begin{equation}\label{f_1}
f_1(t)=-\frac{v_3(t)f_0}{w_3(p,q,t)-z}.
\end{equation}

Substituting the expression (\ref{f_1}) for $f_1$ into the first
equation of the system (\ref{sistem h_3}), we get $f_0
\Delta_3(p,q\,; z)=0.$ If $f_0=0,$ then  $f_1(q)=0.$ This
contradicts the fact that $f=(f_0,f_1)$ is an eigenvector the
operator $h_3(p,q)$. Thus, $\Delta_3(p,q\,; z)=0.$

"If Part." Let for some $z \in {\bf C}\setminus \sigma_{ess}
(h_3(p,q))$ the equality $\Delta_3(p,q\,; z)=0$ hold. It is easy
to show that the vector-function $f=(f_0,f_1)\in {\cal H}^{(0,1)}$
is an eigenvector of the operator $h_3(p,q),\,p,q \in {\bf T}^\nu$
corresponding to the eigenvalue $z \in {\bf C}\setminus
\sigma_{ess} (h_3(p,q)),$ where $f_0=const \neq 0$ and $f_1$ is
defined by (\ref{f_1}). $\Box$

From the Lemma \ref{Delta_3} immediately follows the following
equality
\begin{equation}\label{disc h_3}
\sigma_{disc}(h_3(p,q))=\{z\in {\bf C}\setminus
\sigma_{ess}(h_3(p,q)): \Delta_3(p,q\,; z) =0\},\,p,q\in {\bf
T}^\nu.
\end{equation}

Using definitions of the operators $h_1(p),\,p\in {\bf T}^\nu$ and
$h_2(p),\,p\in {\bf T}^\nu$ we obtain that for any $p\in {\bf
T}^\nu$ the equality $\sigma_{ess}(h_1(p))=\sigma_{ess}(h_2(p))$
holds.

For any fixing $p\in {\bf T}^\nu$ we define an analytic function
$\Delta_1(p\,; \cdot)$ resp. $\Delta_2(p\,; z)$ (the Fredholm
determinant associated with the operator $h_1(p),\,p\in {\bf
T}^\nu$ resp. $h_2(p),\,p\in {\bf T}^\nu$) in ${\bf C}\setminus
\sigma_{ess} (h_1(p))$ by
\begin{equation*}
\Delta_1(p\,; z)=1-\int\limits_{{\bf T}^\nu}
\frac{v_{21}^2(s)ds}{\Delta_3(p,s;z)}
\end{equation*}
resp.
\begin{equation*}
\Delta_2(p\,; z)=\left(1-\int\limits_{{\bf T}^\nu}
\frac{v_{22}^2(s)ds}{\Delta_3(p,s;z)}\right)
\left(w_1(p)-z-\int\limits_{{\bf T}^\nu}\frac{
v_2^2(s)ds}{\Delta_3(p,s;z)} \right)-\left(\int\limits_{{\bf
T}^\nu} \frac{v_2(s)v_{22}(s)ds} {\Delta_3(p,s;z)}\right)^2.
\end{equation*}

Analogously to (\ref{disc h_3}) one can derive the equalities
\begin{equation}\label{disc h_1}
\sigma_{disc}(h_1(p))=\{z\in {\bf C}\setminus
\sigma_{ess}(h_1(p)): \Delta_1(p\,; z) =0\},\,p\in {\bf T}^\nu
\end{equation}
and
\begin{equation}\label{disc h_2}
\sigma_{disc}(h_2(p))=\{z\in {\bf C}\setminus
\sigma_{ess}(h_2(p)): \Delta_2(p\,; z) =0\},\,p\in {\bf T}^\nu.
\end{equation}

{\bf Proof of Theorem \ref{chanl spectr}.} The assertions of the
Theorem \ref{chanl spectr} follows from the representations
(\ref{decom H_3}), (\ref{decom H_i}) and the theorem on
decomposable operators (see \cite{RS-4}) and the equalities
(\ref{disc h_3})-(\ref{disc h_2}). $\Box$

\begin{cor}\label{H_3 subset H_1 H_2} The following inclusion
\begin{equation*}
\sigma(H_3)\subset \sigma(H_1)\cup \sigma(H_2)
\end{equation*}
holds.
\end{cor}

The proof of the Corollary \ref{H_3 subset H_1 H_2} immediately
follows from the Theorem \ref{chanl spectr}.

\section{THE FADDEEV-YAKUBOVSKII TYPE  SYSTEM OF INTEGRAL EQUATIONS AND THE OPERATOR $T(z)$}

In this section we derive an analog of the Faddeev-Yakubovskii
type system of integral equations for the eigenvectors,
corresponding to the eigenvalues lying outside of the essential
spectrum of the operator $H.$

Let us introduce the notations
\begin{equation*}
\overline{\cal{H}}_0={\cal H}_0,\,
\overline{\cal{H}}_1=\overline{\cal{H}}_2=\overline{\cal{H}}_3=
{\cal H}_1 \quad \mbox{and} \quad \overline{\cal{H}}=
\bigoplus\limits_{i=0}^3 \overline{\cal{H}}_i.
\end{equation*}

For each $z\in {\bf C}\setminus (\sigma(H_1) \cup \sigma(H_2) \cup
\sigma(H_3))$ let the operator matrices $A(z)$ and  $K(z)$ act in
the Hilbert space $\overline{\cal{H}}$ as
\begin{equation*}
A(z)=\left( \begin{array}{cccc}
A_{00}(z) & 0 & 0 & 0\\
0 & A_{11}(z) & 0 & A_{13}(z)\\
0 & 0 & A_{22}(z) & 0\\
0 & A_{31}(z) & 0 & A_{33}(z)\\
\end{array}
\right),
\end{equation*}
\begin{equation*}
K(z)=\left( \begin{array}{cccc}
K_{00}(z) & K_{01}(z) & 0 & 0\\
K_{10}(z) & 0 & K_{12}(z) & 0\\
0 & K_{21}(z) & 0 & K_{23}(z)\\
0 & 0 & K_{32}(z) & 0\\
\end{array}
\right),
\end{equation*}
where $A_{ij}(z):\overline{\cal{H}}_j \to \overline{\cal{H}}_i
,\,\ i,j=0,1,2,3$ is the multiplication operator by the function
$a_{ij}(p\,; z)$:
\begin{equation*}
a_{00}(p\,; z)\equiv 1, \quad a_{11}(p\,;
z)=w_1(p)-z-\int\limits_{{\bf
T}^\nu}\frac{v_2^2(s)ds}{\Delta_3(p,s\,;z)},
\end{equation*}
\begin{equation*}
 a_{13}(p\,; z)\equiv a_{31}(p\,; z)=\int\limits_{{\bf
T}^\nu}\frac{v_2(s)v_{22}(s)ds}{\Delta_3(p,s\,;z)},
\end{equation*}
\begin{equation*}
a_{22}(p\,; z)= 1-\int\limits_{{\bf
T}^\nu}\frac{v_{21}^2(s)ds}{\Delta_3(s,p\,;z)},\quad a_{33}(p\,;
z)= 1-\int\limits_{{\bf
T}^\nu}\frac{v_{22}^2(s)ds}{\Delta_3(p,s\,;z)},
\end{equation*}
and the operators $K_{ij}(z): \overline{\cal{H}}_j \to
\overline{\cal{H}}_i ,\,\, i,j=0,1,2,3 $ are defined as
\begin{equation*}
(K_{00}(z)\psi_0)_0=(w_0-z+1)\psi_0, \, K_{01}(z)\equiv H_{01},\,
 K_{10}(z)\equiv -H_{10},
\end{equation*}
\begin{equation*}
(K_{12}(z)\psi_2)_1(p)=-v_{21}(p)\int\limits_{{\bf
T}^\nu}\frac{v_2(s)\psi_2(s)ds}{\Delta_3(p,s\,;z)},
\end{equation*}
\begin{equation*}
(K_{21}(z)\psi_1)_2(p)=-v_2(p)\int\limits_{{\bf
T}^\nu}\frac{v_{21}(s)\psi_1(s)ds}{\Delta_3(s,p\,;z)},
\end{equation*}
\begin{equation*}
(K_{23}(z)\psi_3)_2(p)=v_{22}(p)\int\limits_{{\bf
T}^\nu}\frac{v_{21}(s)\psi_3(s)ds}{\Delta_3(s,p\,;z)},
\end{equation*}
\begin{equation*}
(K_{32}(z)\psi_2)_3(p)=v_{21}(p)\int\limits_{{\bf
T}^\nu}\frac{v_{22}(s)\psi_2(s)ds}{\Delta_3(p,s\,;z)}.
\end{equation*}

We note that for each $z\in {\bf C}\setminus (\sigma(H_1) \cup
\sigma(H_2) \cup \sigma(H_3))$ the operators $K_{ij}(z),\,
i,j=0,1,2$ belong to the Hilbert-Schmidt class and therefore
$K(z)$ is a compact operator.

\begin{lem}\label{A(z) invertible} For each $z\in z\in {\bf C}\setminus (\sigma(H_1)
\cup \sigma(H_2) \cup \sigma(H_3))$ the operator $A(z)$ is bounded
and invertible and the inverse operator $A^{-1}(z)$ is given by
\begin{equation*}
A^{-1}(z)=\left( \begin{array}{cccc}
B_{00}(z) & 0 & 0 & 0\\
0 & B_{11}(z) & 0 & B_{13}(z)\\
0 & 0 & B_{22}(z)  & 0\\
0 & B_{31}(z) & 0  & B_{33}(z)\\
\end{array}
\right),
\end{equation*}
where $B_{ij}(z):\overline{\cal{H}}_j \to \overline{\cal{H}}_i
,\,\ i,j=0,1,2,3$ is the multiplication operator by the function
$b_{ij}(p\,; z)$:
\begin{equation*}
b_{00}(p\,; z)\equiv 1,\quad b_{11}(p\,; z)=\frac{a_{33}(p\,;
z)}{\Delta_2(p\,; z)},\quad b_{13}(p\,; z)=b_{31}(p\,;
z)=-\frac{a_{13}(p\,; z)}{\Delta_2(p\,; z)},
\end{equation*}
\begin{equation*}
b_{22}(p\,; z)=\frac{1}{\Delta_1(p\,; z)}, \quad b_{33}(p\,;
z)=\frac{a_{11}(p\,; z)}{\Delta_2(p\,; z)}.
\end{equation*}
\end{lem}

{\bf Proof.} By the definition $A(z)$ is the multiplication
operator by the matrix $A(p\,; z),$ where
\begin{equation*}
A(p\,; z)=\left( \begin{array}{cccc}
a_{00}(p\,; z) & 0 & 0 & 0\\
0 & a_{11}(p\,; z) & 0 & a_{13}(p\,; z)\\
0 & 0 & a_{22}(p\,; z) & 0\\
0 & a_{31}(p\,; z) & 0 & a_{33}(p\,; z)\\
\end{array}
\right).
\end{equation*}

Obviously, for each $z\in {\bf C}\setminus (\sigma(H_1) \cup
\sigma(H_2) \cup \sigma(H_3))$ the matrix-valued function
$A(\cdot\,; z)$ is a continuous on ${\bf T}^\nu.$ This implies
that $A(z)$ is bounded. Since $\det A(p\,; z)=\Delta_1(p\,; z)
\Delta_2(p\,; z)$ and $z\not\in \sigma(H_1) \cup \sigma(H_2) \cup
\sigma(H_3),$ we have that $\det A(p\,; z))\not=0.$ Therefore for
each $p\in {\bf T}^\nu$ and $z\in {\bf C}\setminus (\sigma(H_1)
\cup \sigma(H_2) \cup \sigma(H_3))$ the matrix $A(p\,; z)$ is
invertible and its inverse matrix has the form
\begin{equation*}
A^{-1}(p\,; z)=\left( \begin{array}{cccc}
a_{00}(p\,; z) & 0 & 0 & 0\\
0 & \frac{a_{33}(p\,; z)}{\Delta_2(p\,; z)} & 0 & -\frac{a_{13}(p\,; z)}{\Delta_2(p\,; z)}\\
0 & 0 & \frac{1}{\Delta_1(p\,; z)} &  0 \\
0 & -\frac{a_{13}(p\,; z)}{\Delta_2(p\,; z)} & 0 & -\frac{a_{13}(p\,; z)}{\Delta_2(p\,; z)}\\
\end{array}
\right).
\end{equation*}

Then for each $z\in {\bf C}\setminus (\sigma(H_1) \cup \sigma(H_2)
\cup \sigma(H_3))$ the matrix-valued function $A^{-1}(\cdot\,; z)$
is a continuous on ${\bf T}^\nu.$ Let $A^{-1}(z)$ be the
multiplication operator by the matrix $A^{-1}(p\,; z)$ acting in
$\overline{\cal{H}}.$ It is easy to show that $A^{-1}(z)$ is the
inverse of $A(z).$ $\Box$

Since for each $z\in {\bf C}\setminus (\sigma(H_1) \cup
\sigma(H_2) \cup \sigma(H_3))$ the operator $A(z)$ is invertible,
for such $z$ we can define the operator $T(z)=A^{-1}(z)K(z).$

The following lemma established a connection between of
eigenvalues of $H$ and $T(z).$

\begin{lem}\label{T(z)} The number $z\in {\bf C}\setminus
(\sigma(H_1) \cup \sigma(H_2) \cup \sigma(H_3))$ is an eigenvalue
of the operator $H$ if and only if the number $\lambda =1$ is an
eigenvalue the operator $T(z).$
\end{lem}

{\bf Proof.} Let $ z\in {\bf C} \setminus (\sigma(H_1) \cup
\sigma(H_2) \cup \sigma(H_3))$ be an eigenvalue of the operator
$H$ and $f=(f_0,f_1,f_2,f_3)\in {\cal H}^{(0,3)}$ be the
corresponding eigenvector, that is, the equation $Hf=zf$ or the
system of equations
\begin{equation*}
((H_{00}-zI_0)f_0)_0+ (H_{01}f_1)_0=0;
\end{equation*}
\begin{equation}\label{sistem equations}
(H_{10}f_0)_1(p)+((H_{11}-zI_1)f_1)_1(p)+(H_{12}f_2)_1(p)=0;
\end{equation}
\begin{equation*}
(H_{21}f_1)_2(p,q)+((H_{22}-zI_2)f_2)_2(p,q)+(H_{23}f_3)_2(p,q)=0;
\end{equation*}
\begin{equation*}
(H_{32}f_2)_3(p,q,t)+((H_{33}-zI_3)f_3)_3(p,q,t)=0
\end{equation*}
has a nontrivial solution $f=(f_0,f_1,f_2,f_3)\in {\cal
H}^{(0,3)},$ where $I_i,\,i=\overline{0,3}$ is an identity
operator in ${\cal H}_i,\,i=\overline{0,3}.$ Since $z\not\in
\sigma_{four}(H_3),$ from the fourth equation of the system
(\ref{sistem equations}) for $f_3$ we have
\begin{equation}\label{f_3}
f_3(p,q,t)=-\frac{v_3(t)f_2(p,q)}{w_3(p,q,t)-z}.
\end{equation}
Substituting the expression (\ref{f_3}) for $f_3$ into the third
equation of the system (\ref{sistem equations}) we obtain that the
system of equations
\begin{equation*}
((H_{00}-zI_0)f_0)_0+ (H_{01}f_1)_0=0;
\end{equation*}
\begin{equation}\label{sistem equations-1}
(H_{10}f_0)_1(p)+((H_{11}-zI_1)f_1)_1(p)+(H_{12}f_2)_1(p)=0;
\end{equation}
\begin{equation*}
(H_{21}f_1)_2(p,q)+((H_{22}-zI_2-H_{23}R_{33}(z)H_{32})f_2)_2(p,q)=0
\end{equation*}
has a nontrivial solution if and only if the system of equations
(\ref{sistem equations}) has a nontrivial solution, where
$R_{33}(z)$ is the resolvent of $H_{33}.$

Since $z\not\in \sigma_{three}(H_3)$ from the third equation of
system (\ref{sistem equations-1}) for $f_2$ we have
\begin{equation}\label{f_2}
f_2(p,q)=-\frac{v_2(q)f_1(p)}{\Delta_3(p,q\,;z)}+
\frac{v_{21}(p)c_1(q)+v_{22}(q)c_2(p)}{\Delta_3(p,q\,;z)},
\end{equation}
where
\begin{equation}\label{c_1}
c_1(q)=\int\limits_{{\bf T}^\nu} v_{21}(s)f_2(s,q)ds,
\end{equation}
\begin{equation}\label{c_2}
c_2(p)=\int\limits_{{\bf T}^\nu} v_{22}(s)f_2(p,s)ds.
\end{equation}

Next we transform the system (\ref{sistem equations-1}) using
$f_0,\,f_1,\,c_1,\,c_2.$ Substituting the expression (\ref{f_2})
for $f_2$ into  the second equation of the system (\ref{sistem
equations-1}) and the equalities (\ref{c_1}), (\ref{c_2}) we
obtain that the system of equations
\begin{equation*}
f_0=(w_0-z+1)f_0 + \int\limits_{{\bf T}^\nu} v_1(s)f_1(s) ds;
\end{equation*}
\begin{equation*}
\left( w_1(p)-z-\int\limits_{{\bf
T}^\nu}\frac{v_2^2(s)ds}{\Delta_3(p,s\,; z)}\right)
f_1(p)+\int\limits_{{\bf T}^\nu}
\frac{v_2(s)v_{22}(s)ds}{\Delta_3(p,s\,; z)}c_2(p)=
\end{equation*}
\begin{equation}\label{sistem of equaton-2}
-v_{1}(p)f_0-v_{21}(p)\int\limits_{{\bf T}^\nu}
\frac{v_{2}(s)c_1(s)ds}{\Delta_3(p,s\,; z)};
\end{equation}
\begin{equation*}
\left( 1-\int\limits_{{\bf
T}^\nu}\frac{v_{21}^2(s)ds}{\Delta_3(s,q\,; z)}\right)
c_1(q)=-v_2(q)\int\limits_{{\bf T}^\nu}
\frac{v_{21}(s)f_{1}(s)ds}{\Delta_3(s,q\,;
z)}+v_{22}(q)\int\limits_{{\bf T}^\nu}
\frac{v_{21}(s)c_{2}(s)ds}{\Delta_3(s,q\,; z)};
\end{equation*}
\begin{equation*}
\int\limits_{{\bf T}^\nu} \frac{v_2(s)v_{22}(s)ds}{\Delta_3(p,s\,;
z)}f_1(p)+ \left(1-\int\limits_{{\bf
T}^\nu}\frac{v_{22}^2(s)ds}{\Delta_3(p,s\,; z)}\right)c_2(p)=
v_{21}(p)\int\limits_{{\bf T}^\nu}\frac{v_{22}(s)c_{1}(s)ds}
{\Delta_3(p,s\,; z)}
\end{equation*}
or the equation
\begin{equation*}
A(z)\psi=K(z)\psi, \,\ \psi=(f_0,f_1,c_1,c_2)\in
\overline{\cal{H}}
\end{equation*}
has a nontrivial solution if and only if the system of equations
(\ref{sistem equations-1}) has a nontrivial solution.

By the Lemma \ref{A(z) invertible} for each $z\in {\bf C}\setminus
(\sigma(H_1) \cup \sigma(H_2) \cup \sigma(H_3))$ the operator
$A(z)$ is invertible and hence the equation
\begin{equation*}
\psi=A^{-1}(z)K(z)\psi
\end{equation*}
or
\begin{equation*}
\psi=T(z)\psi
\end{equation*}
has a nontrivial solution if and only if the system of equations
(\ref{sistem of equaton-2}) has a nontrivial solution. $\Box$

\begin{rem} We point out that the equation $T(z)g=g$ is an analogues of
the Faddeev-Yakubovskii type system of integral equations for
eigenvectors of the operator $H.$
\end{rem}

\section{THE PROOF OF THE MAIN RESULTS}

In this section applying the Weyl criterion and the
Faddeev-Yakubovskii type system of integral equations we prove
Theorem \ref{ess of H}, then the proof of Theorem \ref{HWZ} will
be follow from Theorems \ref{ess of H} and \ref{chanl spectr}.

{\bf Proof of Theorem \ref{ess of H}.} The inclusion $\sigma(H_3)
\subset \sigma_{ess}(H)$ can be proven quite similarly to the
corresponding inclusion of \cite{LR-1}. We prove that
$\sigma(H_1)\cup \sigma(H_2) \subset \sigma_{ess}(H).$

The set $\sigma(H_1)\cup \sigma(H_2)$ we rewrite in the form
\begin{equation*}
\sigma(H_1)\cup \sigma(H_2)=\sigma_{two}(H_1)\cup
\sigma_{two}(H_2) \cup \sigma(H_3).
\end{equation*}

Let $z_0$ be an arbitrary point of $\sigma(H_1)\cup \sigma(H_2).$
There are two cases possible:
\begin{equation*}
1)\,z_0\in \sigma(H_3),
\end{equation*}
\begin{equation*}
2)\, z_0\not\in \sigma(H_3).
\end{equation*}

If $z_0\in \sigma(H_3),$ then $z_0\in \sigma_{ess}(H).$ Let
$z_0\in (\sigma_{two}(H_1)\cup \sigma_{two}(H_2))\setminus
\sigma(H_3).$ By the definition of $\sigma_{two}(H_1)\cup
\sigma_{two}(H_2),$ there exists a point $ p_0 \in {\bf T}^\nu $
such that $\Delta_1(p_0\,; z_0)\Delta_2(p_0\,; z_0)=0.$ Then the
system of homogenous linear equations
\begin{equation*}
l_0=0;
\end{equation*}
\begin{equation*}
\left( w_1(p_0)-z_0-\int\limits_{{\bf
T}^\nu}\frac{v_2^2(s)ds}{\Delta_3(p_0,s\,; z_0)}\right)
l_1+\int\limits_{{\bf T}^\nu}
\frac{v_2(s)v_{22}(s)ds}{\Delta_3(p_0,s\,; z_0)}l_3=0;
\end{equation*}
\begin{equation}\label{sistem linear equat}
\left( 1-\int\limits_{{\bf T}^\nu}\frac{v_{21}^2(s)ds}
{\Delta_3(p_0,s\,; z_0)}\right) l_2=0;
\end{equation}
\begin{equation*}
\int\limits_{{\bf T}^\nu} \frac{v_2(s)v_{22}(s)ds}
{\Delta_3(p_0,s\,; z_0)}l_1+ \left(1-\int\limits_{{\bf
T}^\nu}\frac{v_{22}^2(s)ds}{\Delta_3(p_0,s\,; z_0)}\right)l_3=0
\end{equation*}
has an infinite number of solutions on ${\bf C}^4,$ where ${\bf
C}^4$ is the Cartesian fourth power of ${\bf C}.$ It is easy to
verify that there exists a nontrivial solution ${\bf
l}=(0,l_1,l_2,l_3)\in {\bf C}^4$ of system of equations
(\ref{sistem linear equat}) satisfying one of the following
conditions:

1. If $\Delta_2(p_0\,; z_0)=0,$ then either $l_1\not=0$ and
$l_2=0$ or $l_1=0,\,l_2=0$ and $l_3\not=0.$

2. If $\Delta_1(p_0\,; z_0)=0,$ then $l_2\not=0$ and $l_1=l_3=0.$

The system of equations (\ref{sistem linear equat}) can be written
in the form
\begin{equation*}
A(p_0\,; z_0)l=0, \quad l=(0,l_1,l_2,l_3)\in {\bf C}^4.
\end{equation*}

Let $\chi_{V_n}(\cdot)$ be the characteristic function of the set
\begin{equation*}
V_n(p_0)= \left \{p\in {\bf T}^\nu: \frac{1}{n+1}<|p-p_0|<
\frac{1}{n} \right \},\,n=1,2,\cdots
\end{equation*}
and $\mu(V_n(p_0))$ be the Lebesgue measure of the  set
$V_n(p_0).$

We choose a sequence of orthogonal vector-functions $\{f^{(n)}\}$
as
\begin{equation*}
f^{(n)} = \left ( \begin{array}{cccc}
0 \\
f_1^{(n)}(p)\\
f_2^{(n)}(p, q)\\
f_3^{(n)}(p, q,t)
\end{array}
\right) ,
\end{equation*}
where
\begin{equation*}
f_1^{(n)}(p)=\psi_1^{(n)}(p), \quad p \in {\bf T}^\nu,
\end{equation*}
\begin{equation*}
f_2^{(n)}(p,q)=-\frac{v_2(q)\psi_1^{(n)}(p)}{\Delta_3(p,q\,; z_0)}
+\frac{v_{21}(p)\psi_2^{(n)}(q)+v_{22}(q)\psi_3^{(n)}(p)}{\Delta_3(p,q\,;
z_0)} ,\quad p,q \in {\bf T}^\nu,
\end{equation*}
\begin{equation*}
f_3^{(n)}(p,q,t)=-\frac{v_3(t)f_2^{(n)}(p,q)}{w_3(p,q,t)-
z_0},\quad p,q,t \in {\bf T}^\nu,
\end{equation*}
\begin{equation*}
\psi_i^{(n)}(p)=l_i k_n(p) \chi_{V_n}(p)(\mu(V_n(p_0)))^{-1/2}
,\quad i=1,2,3.
\end{equation*}
Here $ \{k_n\} \subset L_2({\bf T}^\nu)$ is to found from the
orthogonality condition for $ \{f^{(n)} \}, $ i. e.,
\begin{equation*}
(f^{(n)}, f^{(m)}) = \frac{l_2}{\sqrt{\mu (V_n(p_0))} \sqrt{\mu
(V_m(p_0))}} \int \limits_{V_n(p_0)} \int \limits_{V_m(p_0)}
\left(1+\int\limits_{{\bf
T}^\nu}\frac{v_3^2(t)dt}{(w_3(p,q,t)-z_0)^2} \right) \times
\end{equation*}
\begin{equation*}
\left[\frac{v_{21}(p)(l_3v_{22}(q)-l_1 v_2(q))}{\Delta_3^2(p,q\,;
z_0)}+\frac{v_{21}(q)(l_3v_{22}(p)-l_1 v_2(p))}{\Delta_3^2(p,q\,;
z_0)}\right]k_n(p)k_m(q)dpdq=0, \, n\not= m.
\end{equation*}
The existence of $k_n(p)$  follows from the following proposition.

\begin{prop}\label{about k_n} There exists an orthonormal system $\{k_n \} \subset
L_2({\bf T}^\nu),$ satisfying the conditions
\begin{equation*}
a)\, supp \ k_n \subset V_n(p_0) ,
\end{equation*}
\begin{equation*}
b) \int \limits_{V_n(p_0)} \int \limits_{V_m(p_0)}
\left(1+\int\limits_{{\bf T}^\nu}\frac{v_3^2(t)dt}
{(w_3(p,q,t)-z_0)^2} \right) \times
\end{equation*}
\begin{equation*}
\left[\frac{v_{21}(p)(l_3v_{22}(q)-l_1 v_2(q))}{\Delta_3^2(p,q\,;
z_0)}+\frac{v_{21}(q)(l_3v_{22}(p)-l_1 v_2(p))}{\Delta_3^2(p,q\,;
z_0)}\right]k_n(p)k_m(q)dpdq=0, \, n\not= m.
\end{equation*}
\end{prop}

{\bf Proof.} We construct the sequence $ \{ k_n \} $ by induction.
Let
$$ k_1(p)= \chi_{V_1}(p) \left( \sqrt{\mu(V_1(p_0))}
\right)^{-1}.
$$
We choose the function $\tilde{k}_2 \in L_2(V_2(p_0)),$ such that
$ \| \tilde{k}_2 \|_{{\cal H}_1} =1 $ and
$(\tilde{k}_2,\varepsilon_1^{(2)})=0 $ , where
\begin{equation*}
\varepsilon_1^{(2)}(p)= \chi_{V_2}(p) \int\limits_{{\bf T}^\nu}
\left(1+\int\limits_{{\bf T}^\nu}\frac{v_3^2(t)dt}
{(w_3(p,q,t)-z_0)^2} \right) \times
\end{equation*}
\begin{equation*}
\left[\frac{v_{21}(p)(l_3v_{22}(q)-l_1 v_2(q))}{\Delta_3^2(p,q\,;
z_0)}+\frac{v_{21}(q)(l_3v_{22}(p)-l_1 v_2(p))}{\Delta_3^2(p,q\,;
z_0)}\right]k_1(q) dq.
\end{equation*}

We set $ k_2(p)=\tilde{k}_2(p) \chi_{V_2}(p)$ and continue the
process. Assuming that the functions $
k_1(\cdot),\cdots,k_n(\cdot)$ are constructed, we choose the
function $\tilde{k}_{n+1}\in L_2(V_{n+1}(p_0)),$ such that
$\|\tilde{k}_{n+1} \|_{{\cal H}_1}=1$ and it is orthogonal to the
functions
\begin{equation*}
\varepsilon_i^{(n+1)}(p)= \chi_{V_{n+1}}(p) \int\limits_{{\bf
T}^\nu} \left(1+\int\limits_{{\bf T}^\nu}\frac{v_3^2(t)dt}
{(w_3(p,q,t)-z_0)^2} \right) \times
\end{equation*}
\begin{equation*}
\left[\frac{v_{21}(p)(l_3v_{22}(q)-l_1 v_2(q))}{\Delta_3^2(p,q\,;
z_0)}+\frac{v_{21}(q)(l_3v_{22}(p)-l_1 v_2(p))}{\Delta_3^2(p,q\,;
z_0)}\right]k_i(q) dq, \quad i=\overline{1,n} .
\end{equation*}
We set $k_{n+1}(p)= \tilde{k}_{n+1}(p) \chi_{V_{n+1}}(p).$ We have
thus constructed an orthonormalized system $ \{k_n \} $ satisfying
the conditions of the proposition. The Proposition \ref{about k_n}
is proved. $\Box$

{\bf We resume the proof of Theorem \ref{ess of H}.}

We assume that $\Delta_2(p_0\,; z_0)=0$ and $l_1\not=0,\,l_2=0.$
Then
\begin{equation*}
\|f^{(n)}\|_{{\cal H}^{(0,3)}}^2\geq \|f_1^{(n)}\|_{{\cal
H}_1}^2=\frac{d_1}{\mu(V_n(p_0))}, \,\ d_1=l_1^2>0.
\end{equation*}

Let $\Delta_2(p_0\,; z_0)=0$ and $l_3\not=0,\,l_1=l_2=0.$ Then
\begin{equation*}
\|f^{(n)}\|_{{\cal H}^{(0,3)}}^2 \geq \|f_2^{(n)}\|_{{\cal
H}_2}^2= \frac{l_3^2}{\mu(V_n(p_0))} \int\limits_{V_n(p_0)}
\int\limits_{{\bf T}^\nu} \left| \frac{v_{22}(q)k_n(p)}
{\Delta_3(p,q\,; z_0)}\right|^2dpdq \geq
 \frac{d_2}{\mu(V_n(p_0))},
\end{equation*}
\begin{equation*}
 d_2=\frac{l_3^2
\|v_{22}\|_{{\cal H}_1}^2}{\max\limits_{p,q\in {\bf
T}^\nu}|\Delta_3(p,q\,; z_0)|^2}.
\end{equation*}

Similarly, we can prove that in the case where $\Delta_1(p_0\,;
z_0)=0,\, l_2\not=0$ and $l_1=l_3=0$ the inequality
\begin{equation*}
\|f^{(n)}\|_{{\cal H}^{(0,3)}}^2 \geq \frac{d_3}{\mu(V_n(p_0))},\,
d_3=\frac{l_2^2 \|v_{21}\|_{{\cal H}_1}^2}{\max\limits_{p,q\in
{\bf T}^\nu}|\Delta_3(p,q\,; z_0)|^2}
\end{equation*}
holds. Therefore
\begin{equation}\label{norm of f_n}
\|f^{(n)}\|_{{\cal H}^{(0,3)}}^2\geq \frac{\xi_0}{\mu(V_n(p_0))},
\end{equation}
where
\begin{equation*}
 \xi_0=\min\{d_1, d_2, d_3 \}>0.
\end{equation*}

We set $\widetilde{f}^{(n)}=f^{(n)}/ \|f^{(n)}\|_{{\cal
H}^{(0,3)}}.$ It is clear that the system
$\{\widetilde{f}^{(n)}\}$ is orthonormal.

We consider the operator $(H-z_0)\widetilde{f}^{(n)}$ and estimate
its norm as
\begin{equation*}
\| (H-z_0)\widetilde{f}^{(n)} \|_{{\cal H}^{(0,3)}} \leq
\|A(z_0)\widetilde{\psi}^{(n)}\|_{\overline{\mathcal{H}}} +
\|K(z_0)\widetilde{\psi}^{(n)}\|_{\overline{\mathcal{H}}},
\end{equation*}
where
\begin{equation*}
\widetilde{\psi}^{(n)}=\left(0,\frac{\psi^{(n)}_1}{\|f^{(n)}\|_{{\cal
H}^{(0,3)}}}, \frac{\psi^{(n)}_2}{\|f^{(n)}\|_{{\cal H}^{(0,3)}}},
\frac{\psi^{(n)}_3}{\|f^{(n)}\|_{{\cal H}^{(0,3)}}}\right).
\end{equation*}

We note that  $\{ \widetilde {\psi}^{(n)}\}\subset
\overline{\mathcal{H}}$ is bounded orthonormal system. Indeed, the
othogonality follows since for any $n\neq m$ the supports of the
functions $\widetilde{\psi}^{(n)}$ and $\widetilde{\psi}^{(m)}$
are nonintersecting. The equality
\begin{equation*}
\|\widetilde{\psi}^{(n)}\|_{\overline{\cal H}}^2=
\frac{1}{\|f^{(n)}\|_{{\cal
H}^{(0,3)}}^2}\frac{1}{\mu(V_n(p_0))}(l_1^2+l_2^2+l_3^2)
\end{equation*}
and inequality (\ref{norm of f_n}) imply that the system
$\{\widetilde{\psi}^{(n)}\}$ is uniformly bounded, i. e.,
\begin{equation*}
\|\widetilde{\psi}^{(n)}\|_{\overline{\cal H}}\leq \frac{1}{\xi_0}
\left\|\textbf{l}\right\|^2_{{\bf C}^4}
\end{equation*}
for all positive integers $n.$

Since the operator $K(z_0)$ is a compact, it follows that $
\|K(z_0)\widetilde{\psi}^{(n)} \|_{\overline{\cal H}}\rightarrow
0$ as $n\rightarrow\infty.$

We next  estimate
$\|A(z_0)\widetilde{\psi}^{(n)}\|_{\overline{\cal H}}.$  Applying
the Schwarz inequality we have
\begin{equation*}
\| A(z_0)\widetilde{\psi}^{(n)} \|_{\overline{\cal H}}^2 \leq M^2
\sup \limits_{p\in {V_n(p_0)}}\left\|A(p\,; z_0)\textbf{l}
\right\|_{{\bf C}^4}^2\quad \mbox{with} \quad M^2=\max
\{\frac{2}{\xi_0}, \frac{\|v_{22}\|_{{\cal H}_1}^2}{\xi_0},
\frac{\|v_{21}\|_{{\cal H}_1}^2}{\xi_0}\}.
\end{equation*}

The continuity of the matrix-valued function  $A(\cdot\,; z_0)$
implies that
\begin{equation*}
\sup\limits_{p\in V_n(p_0)}\|A(p\,; z_0)\textbf{l}\|_{{\bf
C}^4}\rightarrow 0 \quad \mbox{as} \quad n\rightarrow \infty.
\end{equation*}

Therefore, for the sequence of orthonormal vector-functions  $ \{
\widetilde{f}^{(n)} \} $ it follows that
$$
\| (H-z_0)\widetilde{f}^{(n)} \|_{{\cal H}^{(0,3)}} \to 0\quad
\mbox{as} \quad n\to\infty
$$
and hence by Weyl's criterion we have that $z_0 \in
\sigma_{ess}(H).$ Since $z_0$ is an arbitrary point of
$\sigma(H_1) \cup \sigma(H_2),$ it follows that $\sigma(H_1)\cup
\sigma(H_2) \subset \sigma_{ess}(H).$ Thus we have proved that
$\sigma(H_1) \cup \sigma(H_2) \cup \sigma(H_3) \subset
\sigma_{ess}(H).$

Now we prove the converse inclusion, that is, $ \sigma_{ess}(H)
\subset \sigma(H_1) \cup \sigma(H_2) \cup \sigma(H_3).$ Since for
any $z\in {\bf C}\backslash (\sigma(H_1) \cup \sigma(H_2) \cup
\sigma(H_3))$ the operator $K(z)$ is a compact and $A^{-1}(z)$ is
bounded, we have that $f(z)=A^{-1}(z)K(z)$ is a compact-valued
analytic function in ${\bf C}\backslash (\sigma(H_1) \cup
\sigma(H_2) \cup \sigma(H_3)).$ From the self-adjointness of $H$
and Lemma \ref{T(z)} it follows that the operator $({\bf
I}-f(z))^{-1}$ exists for all $Im z\not=0,$ where ${\bf I}$ is an
identity operator in $\overline{\cal H}.$ In accordance with the
analytic Fredholm theorem, we conclude that the operator-valued
function $({\bf I}-f(z))^{-1}$ exists on ${\bf C}\setminus
(\sigma(H_1) \cup \sigma(H_2) \cup \sigma(H_3))$ everywhere except
at a discrete set $S,$ where it has finite-rank residues. Hence,
with $\sigma_{disc}(H)$ denoting the discrete spectrum of $H,$ we
have $\sigma(H)\setminus \sigma(H_1) \cup \sigma(H_2) \cup
\sigma(H_3) \subset \sigma_{disc}(H)=\sigma(H)\setminus
\sigma_{ess}(H),$ i. e., $\sigma_{ess}(H)\subset (\sigma(H_1) \cup
\sigma(H_2) \cup \sigma(H_3)).$ The Theorem \ref{ess of H} is
completely proved. $\Box$

The proof of Theorem \ref{HWZ} follows from the Theorems \ref{ess
of H} and \ref{chanl spectr}.

\section{BLOCK OPERATOR MATRICES AND A VARIA\-TIO\-NAL TECHNIQUE}

In this section we give a variational technique to find the
boundaries of $\sigma_{ess}(H)$ and eigenvalues from some interior
part of the discrete spectrum of $\sigma(H).$ Define

$$ A=\left(
\begin{array}{cc}
H_{00}& H_{01}\\
H_{10}&H_{11}\\
\end{array}
 \right),
\quad C=\left(
\begin{array}{cc}
H_{22}& H_{23}\\
H_{32}&H_{33}\\
\end{array}
 \right),
$$
$$
B=\left(
\begin{array}{cc}
0& 0\\
H_{12}&0\\
\end{array}
 \right),
\quad B^*=\left(
\begin{array}{cc}
0& H_{21}\\
0&0\\
\end{array}
 \right),
$$
where $A: \mathcal{H}^{(0,1)}\rightarrow \mathcal{H}^{(0,1)}$,
$C:\mathcal{H}^{(2,3)}\rightarrow \mathcal{H}^{(2,3)},$ $B:
\mathcal{H}^{(2,3)}\rightarrow \mathcal{H}^{(0,1)}$ and
$B^*:\mathcal{H}^{(0,1)}\rightarrow \mathcal{H}^{(2,3)}.$

Then the operator $H$ acting in Hilbert space
$\mathcal{H}^{(0,1)}\bigoplus \mathcal{H}^{(2,3)}$ defined by
(\ref{main operator}) can be written as a symmetric operator
matrix in the form
\begin{equation}\label{second form H}
H=\left(
\begin{array}{cc}
A& B\\
B^*&C\\
\end{array}
 \right).
\end{equation}

For any subspace $S\subset \mathcal{H}^{(0,3)}$ define $S^1=\{x\in
S: \|x\|= 1\}$ and let
\begin{equation*}
\lambda_i(H)= \inf_{S\in S_i}\max_{S^1} (H x,\,x)
\end{equation*}
and
\begin{equation*}
\mu_i(H)= \sup_{S\in S_i}\min_{S^1} (H x,\,x),
\end{equation*}
where $S_i$ denotes the set of all subspaces of dimension $i.$
These numbers, in general are not eigenvalues of $H.$ Now define
the boundaries of the spectrum and essential spectrum of $H$ as
\begin{equation*}
a(H)= \min \sigma(H),\, b(H)= \max \sigma(H),
\end{equation*}
\begin{equation*}
a_{ess}(H)= \min \sigma_{ess}(H),\, b_{ess}(H)= \max
\sigma_{ess}(H).
\end{equation*}

Notice that, although we are mainly interested in spectral
properties for the operator $H,$ but the facts given below valid
for a self-adjoint operator and symmetric operator matrices of the
form (\ref{second form H}).

It is known from the classical Courant-Hilbert-Weyl variational
theory that, if $ a(H)< a_{ess}(H)$ then the spectrum of $H$ in
$[a(H),\,a_{ess}(H))$ is discrete. Moreover, if eigenvalues in
$[a(H),\,a_{ess}(H))$ arranged in increasing order, including
multiplicities, then they are equal to
\begin{equation*}
\lambda_i(H)= \inf_{S\in S_i}\max_{S^1} (H x,\,x),\,\,\,
i=1,2,\cdots,
\end{equation*}
where the $\inf$ is attained, i. e., there exists a subspace $S\in
S_i$ such that
\begin{equation*}
\lambda_i(H)= \max_{S^1} (H x,\,x).
\end{equation*}

Now we give some results from the classical variational theory to
obtain $a_{ess}(H)= \min \sigma_{ess}(H).$ First, if $a_{ess}(H)=
a(H)$ then it means that $ \min \sigma_{ess}(H)= \min\sigma(H)$
and for this reason we let $a(H)<a_{ess}(H).$ Two cases are
possible (see \cite{WS}, Theorem 1, p. 12):

I) The first case: $\lambda_1(H), \lambda_2(H),
\cdots,\lambda_n(H)$ are attained but $\lambda_{n+1}(H)$ is not
attained, i. e.,
\begin{equation*}
\lambda_i(H)= \min_{S\in S_i}\max_{S^1} (H x,\,x),
\end{equation*}
for $i= 1,2,\cdots,n,$
\begin{equation*}
\lambda_{n+1}(H)= \inf_{S\in S_{n+1}}\max_{S^1} (H x,\,x)
\end{equation*}
and no subspace $S\in S_{n+1}$ such that
\begin{equation*}
\lambda_{n+1}(H)= \max_{S^1} (H x,\,x).
\end{equation*}

Then there are only $n$ eigenvalues of $H$ in  $[a(H),\,
a_{ess}(H)),$ which can be described by
\begin{equation}\label{form 6.1}
\lambda_i(H)= \min_{S\in S_i}\max_{S^1} (H x,\,x),\,\,\,
i=1,2,\cdots,n
\end{equation}
 and
\begin{equation}\label{form 6.2}
\min \sigma_{ess}(H):= a_{ess}(H)= \lambda_{n+1}=
\lambda_{n+2}=\cdots.
\end{equation}

2) The second case: all of the numbers $\lambda_i(H)$ are
attained, i. e.,
\begin{equation*}
\lambda_i(H)= \min_{S\in S_i}\max_{S^1} (H x,\,x),\quad i=
1,2,\cdots,n,\cdots.
\end{equation*}
Then in this case the spectrum of $H$ in $[a(H), a_{ess}(H))$
consists of countable number of eigenvalues $\lambda_i(H)$ and
\begin{equation}\label{form 6.3}
a_{ess}(H)= \lim _{n\rightarrow \infty}\lambda_{n}(H).
\end{equation}

The same results hold for $b_{ess}(H)= \max\sigma_{ess}(H)$ but we
need to replace $\lambda_i(H)$ by $\mu_i(H).$

Note that classical variational principles are applicable mainly
to describe the discrete spectrum at the end parts of $\sigma(H).$
Dipper results can be obtained if we use the following operator
function (operator pencil) technique. Here we give a method (see
\cite{BEL,EL-1,EL-2}) which allows to find eigenvalues in some
interior parts of the discrete spectrum of symmetric block
operator matrices (particularly, for $H$) of the form (\ref{second
form H}).

First we give a short information on operator functions (see
\cite{BEL,EL-1,EL-2,Has}). Denote by $S({\cal H})$ the space of
bounded symmetric operators on a Hilbert space ${\cal H}$. Let
$L(\lambda)$ be an operator valued function ( or simply operator
function), defined on an interval $[\alpha, \beta]$ with values in
$S({\cal H})$. So,
\begin{equation*}
L: [\alpha,\,\beta]\rightarrow S({\cal H}).
\end{equation*}
A typical example is an operator polynomial of the form
$L(\lambda)= \lambda^n A_n + \lambda^{(n-1)} A_{n-1}+\ldots+ A_0,$
where $ A_i\in S({\cal H}),\,i=0,1,...,n.$ Polynomial operator
functions are often called operator pencils. We denote the
spectrum and the essential spectrum of $L$ by $\sigma(L)$ and
$\sigma_{ess}(L),$ respectively. Define the family of functions
$\varphi_x(\lambda)$ depending on $x\in {\cal H} \setminus\{0\}$
by
\begin{equation*}
\varphi_x(\lambda):= (L(\lambda)x,\,x).
\end{equation*}
In the variational theory of operator  functions it is always
supposed the following two conditions are satisfied:

\textbf{I)} The equation  $\varphi_x(\lambda)= 0$ has at most one
solution on $[\alpha,\,\beta]$ (which is denoted by $p(x)$) and
$\varphi_x(\lambda)$ is decreasing (or increasing) at $p(x),$ i.
e.,
\begin{equation*}
\varphi_x(\lambda)<0 \Leftrightarrow \lambda > p(x),
\end{equation*}
\begin{equation*}
\varphi_x(\lambda)>0 \Leftrightarrow \lambda < p(x).
\end{equation*}

\textbf{II)} $ \kappa_{\alpha}(L):= \max \dim\{E \Bigr
|\varphi_x(\lambda)< 0,\,\,\, x\in E \setminus\{0\}\}< +\infty.$

By these conditions if the equation $\varphi_x(\lambda)= 0$ has no
solution on $[\alpha, \beta]$ for some $x$ then either
$\varphi_x(\lambda)< 0$ or $\varphi_x(\lambda)> 0$ for all
$\lambda\in [\alpha,\,\beta].$ Clearly, the functional $p(x)$ in
general is not defined for all $x\neq 0$ and in this case we
define the extended functional $p(x)$ as
\begin{equation*}
p(x)=\left\{\begin{array}{ccc} \lambda_0,\,\,&\mbox{
if}\,\,\varphi_x(\lambda_0)= 0,\\+\infty, & \mbox{ if
}\,\,\varphi_x(\lambda)> 0,\,\,\lambda\in [\alpha,\,\beta],
\\-\infty, & \mbox{ if
}\varphi_x(\lambda)< 0,\,\,\lambda\in [\alpha,\,\beta].
\end{array}\right.
\end{equation*}

Now the same formulas (\ref{form 6.1}), (\ref{form 6.2}) and
(\ref{form 6.3}) (under the same conditions) hold (see \cite{BEL},
\cite{Has}) for the operator function $L(\lambda)$ if we replace
$(Hx,\,x)$ by $p(x)$ and define
\begin{equation*}
\lambda_i(L)= \inf_{S\in S_i}\sup\limits_{S^1} p(x)
\end{equation*}
and
\begin{equation*}
\mu_i(L)= \sup_{S\in S_i}\inf_{S^1} p(x).
\end{equation*}
It means that in the spectral theory of operator functions the
functional $p(x),$ which is called a Rayleigh functional, plays
the same role as the quadratic form $\frac{(Hx,\,x)}{(x,\,x)}$ of
the operator $H$ in the operator theory.

Let $I^{(n,m)}$ be an identity operator in ${\cal
H}^{(n,m)},\,0\leq n<m\leq 3.$

Now we give a connection between the spectrum of $H$ of the form
(\ref{second form H}) and the spectrum of the operator function
defined below (see for details \cite{BEL,EL-1,EL-2}). The first
important step is

\begin{thm}\label{6.1} (\cite{BEL,EL-1,EL-2}). The
spectrum of $H$ of the form (\ref{second form H})  outside of the
spectrum $C$ coincides with the spectrum of the operator function
\begin{equation*}
L(\lambda)= A- \lambda I^{(0,1)}- B(C-\lambda I^{(2,3)})^{-1}B^*.
\end{equation*}
\end{thm}

Finally, we give a theorem which shows how one can obtain
eigenvalues from some interior parts of the discrete spectrum of
$H$ by using the Rayleigh functional  $p(x)$ for $L(\lambda)$ (see
\cite{BEL}, pp. 204-205).

\begin{thm}\label{6.2} Let $H$ be an operator matrix of the form (\ref{second form H})
and $\sigma(C)< \sigma_{ess}(A).$ Then,

1) the spectrum of $H$ in $(b(C),\,a_{ess}(A))$ is discrete with
only possible accumulation point at $a_{ess}(A).$

2) For $L(\lambda)= A- \lambda I^{(0,1)}- B(C-\lambda
I^{(2,3)})^{-1}B^*$ we have
\begin{equation*}
\sigma(L)\cap( b(C),\,a_{ess}(A))]= \sigma(H)\cap( b(C),\,
a_{ess}(A))],
\end{equation*}
3) if $\lambda_i(H),\,\, i= 1,2,3,\cdots$ are eigenvalues of $H$
in $( b(C),\,a_{ess}(A)),$ then
\begin{equation*}
\lambda_i(H)= \inf_{S\in S_i}\max_{S^1} p(x),
\end{equation*}
where $p(x)$ is the Rayleigh functional for $L(\lambda).$
\end{thm}

\textbf{A sketch of the proof.} Evidently, it is enough two show
that the operator function $L(\lambda)$ satisfies the condition
\textbf{I)} and \textbf{II)} on  $(b(C),\,a_{ess}(A)).$ It follows
from the Hilbert identity $R_{\lambda}(C)- R_{\mu}(C)= (\lambda-
\mu)R_{\lambda}(C)R_{\mu}(C)$ that $R_{\lambda}^\prime(C)=
R_{\lambda}^2(C),$ where $R_{\lambda}(C):= (C-\lambda
I^{(2,3)})^{-1}$ is the resolvent of the operator $C$. Using this
fact we have
\begin{equation*}
L^\prime(\lambda)= -I^{(1,2)}- B(C-\lambda I^{(2,3)})^{-2}B^*\ll 0
\end{equation*}
for all $\lambda\in ( b(C),\,a_{ess}(A)].$ Here
$L^\prime(\lambda)\ll 0$ means $(L^\prime(\lambda)x,\,x)\leq
\delta (x,\,x)$ for all $x$ and some $\delta > 0.$

In fact by the spectral theorem for a self-adjoint operator we can
write
\begin{equation*}
C= \int\limits_{\sigma(C)}s\,dE_C(s),
\end{equation*}
where $E_C(S)$ is the spectral measure of the operator $C.$ Then
\begin{equation*}
C- \lambda I^{(2,3)}= \int\limits_{\sigma(C)}(s-
\lambda)\,dE_C(s)< 0,
\end{equation*}
because $s-\lambda <0$ for $\lambda > b(C).$  Notice that the
inequality $C-\lambda I^{(2,3)}< 0$ also follows from the fact
that the spectrum of a bounded operator is a subset of the closure
of its numerical range. Now we have $ L^\prime(\lambda)\ll 0$ and
the condition \textbf{I)} follows from this inequality. On the
other hand for $\alpha= b(C)$ we can write
\begin{equation*}
\{x \Bigr | (L(\alpha)x,\,x)< 0\}\subset \{x \Bigr | ((A-\alpha
I^{(0,1)})x,\,x))< 0\}.
\end{equation*}
It follows from this that $\kappa_\alpha(\lambda)\leq
N(\alpha,\,A),$ where $N(\alpha,\,A)$ is the spectral distribution
function of $A$. The condition $\alpha= b(C)< \sigma_{ess}(A)$
means that $N(\alpha,\,A)$ is finite and  by the inequality
$\kappa_\alpha(\lambda)\leq N(\alpha,\,A)$ we get
\begin{equation*}
\kappa_\alpha(\lambda) =\max\{E \Bigr| L(\alpha)x,\,x)<0,\,\, x\in
E \setminus\{0\}<+\infty,
\end{equation*}
i. e., the condition \textbf{II)} is satisfied. Consequently, the
eigenvalues of the operator $H$ in the interval
$(b(C),\,a_{ess}(A))$ can be characterized by variational
principles for the operator function $L(\lambda)$ (see
\cite{BEL,EL-1,EL-2,Has}). More precisely,
$$
\lambda_i(H)= \min_{S\in S_i}\sup\limits_{S^1}
p(x),\,\,i=1,2,\cdots,
$$
where $p(x)$ is the Rayleigh functional of the operator function
$L(\lambda)= A- \lambda I^{(0,1)}- B(C-\lambda I^{(2,3)})^{-1}B^*$
on $[b(C), a_{ess}(A)]$.

\vspace{0.1cm}

{\bf ACKNOWLEDGEMENTS.}  The first author would like to thank the
Abdus Salam International Centre for Theoretical Physics, Trieste,
Italy, for the kind hospitality and support and the Commission on
Development and Exchanges of the International Mathematical Union
for travel grant.

\vspace{0.2cm}

{\sc Tulkin H. Rasulov\\
Samarkand State University, Department of Physics and Mathematics,
15 University Boulevard, Samarkand, 140104, Uzbekistan}\\
E-mail: tulkin$_{-}$rasulov@yahoo.com \vspace{0.2cm}

{\sc Mukhiddin I. Muminov\\
Samarkand State University, Department of Physics and Mathematics,
15 University Boulevard, Samarkand, 140104, Uzbekistan}\\
E-mail: mmuminov@mail.ru \vspace{0.2cm}

{\sc Mahir Hasanov\\
Southern Alberta Institute of Technology, 1301 16th Ave NW
Calgary, Alberta, CANADA T2M 0L4}\\ E-mail: hasanov61@yahoo.com

\end{document}